\documentclass{article}

\usepackage{arxiv}
\usepackage{hyperref}       
\usepackage{url}            
\usepackage{booktabs}       
\usepackage{amsfonts}       
\usepackage{nicefrac}       
\usepackage{microtype}      
\usepackage[utf8]{inputenc}
\usepackage[english]{babel}
\usepackage{amsmath,amsfonts,amssymb}
\usepackage{lmodern}
\usepackage[scaled]{helvet}
\usepackage[T1]{fontenc}
\usepackage{lettrine} 
\usepackage[round]{natbib} 
\usepackage{tabularx}

\usepackage{graphicx}
\usepackage{verbatim}

\title{{\fontsize{16.5}{36pt}\selectfont The Cognitive Science of Extremist Ideologies Online} }

\author{
  Chloe Perry \\
Department of American Culture \\
University of Michigan \\
505 S. State Street \\
Ann Arbor, MI 48109 USA \\ \\
Social \& Decision Sciences \\
Carnegie Mellon University \\ 5000 Forbes Avenue \\
Pittsburgh, PA 15213 USA \\ \\
   \And
 Simon DeDeo \\
Social \& Decision Sciences \\
Carnegie Mellon University \\ 5000 Forbes Avenue \\
Pittsburgh, PA 15213 USA \\ \\
Santa Fe Institute \\
1399 Hyde Park Road \\
Santa Fe, NM 87501 USA \\
  \texttt{sdedeo@andrew.cmu.edu}}

\begin{document}

\maketitle

%

\begin{abstract}
Extremist ideologies are finding new homes in online forums. These serve as both places for true believers, and recruiting-grounds for curious newcomers. To understand how newcomers learn ideology online, we study the Reddit archives of a novel sexist ideology known as the ``the Red Pill''. Matching a longstanding hypothesis in the social sciences, our methods resolve the ideology into two components: a ``behavioral'' dimension, concerned with correcting behavior towards the self and others, and an ``explanatory'' dimension, of unifying explanations for the worldview. We then build a model of how newcomers to the group navigate the underlying conceptual structure. This reveals a large population of ``tourists'', who leave quickly, and a smaller group of ``residents'' who join the group and remain for orders of magnitude longer. Newcomers are attracted by the behavioral component, in the form of self-help topics such as diet, exercise, and addiction. Explanations, however, keep them there, turning tourists into residents. They have powerful effects: explanation adoption can more than double the duration of median engagement, and can explain the emergence of a long-tail of high-power engagers. The most sticky explanations, that predict the longest engagement, are about status hierarchies.
%
\end{abstract}
\vspace{0.5cm}


Extremist ideologies are not new---``Islamism''~\citep{mandaville2010global} or ``white supremacy''~\citep{zanden} are decades or centuries old---but a great deal of contemporary attention has centered on the new role of online forums in how they spread, evolve, and attract adherents. The connection between an online forum and an ideology is often made when a forum participant commits an act of political violence~\citep{nagle2017kill}, but outliers are only one side of the story. Another side is the hundreds of thousands of individuals who encounter these ideologies, experiment with them or even adopt them for a time, but who do not become terrorists themselves. The goal of this paper is to understand the cognitive processes involved in how people learn ideology. 

As we discuss in detail below, all ideologies have a dual nature: they are both explanations of the world, and patterns of behaviors. A sexist ideology, for example, includes not only a network of re-enforcing beliefs that explain a person's experiences in terms of the inferiority of women to men, but also habits of action ranging from degrading comments to physical assault~\citep{manne2017down}.  As explanations, ideologies are of great interest because of how they both link together ideas in ways that appeal to basic sense-making drives. As patterns of behavior, meanwhile, they matter because of how they alter and create basic features of social life---and, of course, because of how damaging the behaviors can become.

Psychological studies of ideology have use survey methods to probe the relationship between these two dimensions. In the psychology of gender, for example, ``traditional masculinity ideology''~\citep{pleck95} is studied using the MRNI~\citep{levant1992male}, a 57-item Likert scale questionnaire that measures the extent to which the subject endorses beliefs that are consequences of different dimensions of the ideology (\emph{e.g.}, the dimension of ``dominance'', which includes agreement to the statement that ``men should be the leader in any group''). A subject's endorsement can then be linked to any number of outcomes on the behavioral side, from aggressive driving~\citep{braly2018threats} to systemic aspects of a person's life such as behavior in intimate relationships, parental engagement, or attitudes towards minority groups (for a recent review, see~\citealp{gerdes2018content}).

By connecting belief and behavior, survey-based research gives us important insights, but it also leaves a number of key questions unanswered. In particular, survey methods have difficulty determining what it was about an ideology that made it initially appealing to someone, and they can not study how an individual came to learn, and adopt, the ideology over time. They also face methodological challenges such as construct validity, because it can take many years to refine a questionnaire, while the ideologies themselves are under constant evolution.

This is particularly challenging when the ideology in question is, as often happens online, a new variant or unexpected combination, rather than a repetition of something already well-understood, and where cultural evolution can act on such a rapid timescale that qualitative construction and iterative refinement can't keep up. Finally, survey-based methods require recruitment of a sufficiently large number of participants who subscribe to the ideology in question, which makes it difficult, if not impossible, to study ideologies when they are still associated with niche subcultures.

In order to address these gaps, we present a new combination of methods (linkage networks combined with hidden Markov Modeling), and apply it to a detailed study of an ideology, known as ``The Red Pill''. The Red Pill appeared in the online forum Reddit between 2010 and 2020, and attracted hundreds of thousands of young men in the U.S. and U.K.. It is a sexist ideology, under the standard definition, anchored in pseudoscience, and is characterized by the dehumanization of women as biological machines, and a call to fight a conspiracy against ``masculine'' values. It has led to, among other things, more than a dozen suicide attacks against perceived supporters of the conspiracy. It is also a ``born-digital'' ideology, meaning that its main features developed over the course of online interactions, rather than a traditional, pre-internet ideology whose adherents migrated online. 

The large scale of our dataset provides a new window onto how this ideology gains traction and mindshare among internet users. It enables us to study the conceptual structure of the ideology in ``high-resolution'', teasing apart different components of the ideology and providing new insight to different variants of ``fellow travelers'' that coexist within what appears to be a homogenous group. The long time-span of the data also provides longitudinal information.  This means that we can study how individuals interact with the ideology over time; we can see where they begin in the conceptual network, which parts of that network are particularly ``sticky'', and where, once caught, the individual tends to go next. This enables us to study the mechanisms of joining, learning, and exit, and to connect them to basic questions in cognitive science such as the role of explanatory values~\citep{WOJTOWICZ2020981} and styles of moral reasoning such as those provided by Moral Foundations Theory~\citep{GRAHAM201355}.

\section{The Cognitive Science of Ideology}

To study the general features of ideology---online or off, extremist or normative---requires clarity about the underlying concept. This is complicated by the fact that the word itself is often used in a pejorative sense: for Marx and Engels, for example, an ideology is an illusion, a falsification of political reality, and those who believe it are fundamentally deceived~\citep{marx, thompson}. By the 1970s, however, sociologists had developed more rhetorically neutral, descriptive meanings; in this second sense, ideologies are simply systems of belief that have to do with political action. While there are a number of different ways this can be fleshed out, a usefully generic definition is that provided by \cite{seliger2019ideology}: ideologies are ``sets of ideas by which [people] posit, explain, and justify the ends and means of organized social action with the aim to preserve, amend, uproot, or rebuild a given reality''. Implicit in this definition is the idea that there are two principal dimensions of ideology; \cite{seliger1970fundamental} calls these the ``fundamental'', and the ``operative''. We take the two in turn.

On the fundamental dimension, an ideology is a cognitive artifact, a connected set of ideas that people use to make sense of the world. Ideologies serve as frameworks that knit statements of fact together with metaphysical, aesthetic, and moral claims, to make a more-or-less coherent and rationally-justified whole. Any particular claim that an ideology makes is meaningful only when understood as part of that larger system. An ideology is not a list of disconnected grievances, but a ``programmatic congerie of ideas''~\citep{friedrich1965ideology} that forms a compelling whole, an abstraction ``less abstract than the abstractions contained within it''~\citep{seliger1979inseparability}. 

Along this dimension, the study of ideology is the psychology of explanation. We value explanations not just for how their component ideas account for the facts at hand (how well they describe the world, or ``descriptiveness''), but also for the ways in which those ideas link the facts together (``co-explanation''), and the ways in which the ideas themselves are linked together (``unification'' and other forms of simplicity; \citealp{WOJTOWICZ2020981}). In a similar fashion, an appealing ideology is, in part, an appealing explanation, one that simplifies the world and helps satisfy the basic psychological drive for sense-making~\citep{CHATER2016137}. For this reason, we refer to the first dimension as \emph{explanatory}. 

On the operative, or behavioral, dimension, an ideology is a political artifact, directed towards the goal of either preserving or transforming a social order. In this sense, ideologies are interpersonal objects, something that we hold in common with other members of a group and that gives us the ability to organize and act together. Laws and social practices may be part of the operative side of an ideology: a racist ideology, for example, may include both laws that discriminate against minorities, as well as oppressive patterns of behavior by members of the majority group in everyday life. These aspects of the operative dimension then accompany the fundamental dimension's explanations for why those behaviors are justified. 

\cite{seliger1970fundamental}'s account emphases the conscious adoption of behaviors, and in as much as the operative dimension of an ideology is strategic, members may wish to seek common knowledge that others share it~\citep{chwe2013rational}. However, as the example of racist ideology suggests, much of what is operative may be part of a group's \emph{habitus}~\citep{bourdieu1990logic}---patterns of action that individuals produce ``without thinking'', and judgements that individuals would say ``go without saying''. The behaviors may be unreflective, but this does not mean they are incoherent, or any less aligned with the ideology's fundamental, explanatory dimension. As pointed out by \cite{converse}, the ordinary citizen has very little conscious awareness of the explanatory frameworks of the political elites, and often relies on simple habits and heuristics, such as identity and judgements of relative power and benefit, to guide the practical behaviors those explanations imply. Indeed, a habitual logic in the behavioral dimension may stabilize before the explanatory dimension does, as happens in \cite{elias2000civilizing}'s account of how the emergence of the modern ideas of the state's monopoly on violence was preceded by shifts in habits of interpersonal etiquette.

The explanatory and behavioral dimension of an ideology are intimately related. The behavioral dimension is often highly structured by the explanatory dimension, and even when an agent experiences his actions as instinctive or ``natural'', they can still be playing out a complex explanatory logic. As pointed out by~\cite{manne2017down}, for example, a man who subscribes to a sexist ideology may not generically ``hate'' women; he may, in fact, display sincere affection to them so long as they fulfill expectations set by the underlying explanations he holds about their proper role in society; laboratory studies can tease apart different kinds of sexist ideology in part by the valence of these responses \citep{glick1997two,fiske}. In a similar fashion, a woman who subscribes to a sexist ideology may gain self-esteem by fulfilling the ideology's criteria for what it means to be a ``good'' woman~\citep{self_esteem}. As suggested by these examples, the behavioral dimension of an ideology is often interpersonal. Ideologies may occasionally involve private actions (\emph{e.g.}, secret acts of ``mortification of the flesh''), but with rare exceptions they center around behaviors towards others, and usually include public acts and political behaviors such as voting.

Once we see ideologies in this way, two basic questions arise. First, what makes an ideology satisfying or appealing? What aspects of the explanatory and behavioral features of an ideology draw people in, and what aspects are ``sticky'', \emph{i.e.}, act to keep them there once they arrive? A simple answer is that people adopt ideologies that make them feel good, but as \cite{CHATER2016137} point out, there are constraints on our thoughts that limit the ways in which we can adopt beliefs for purely hedonic reasons~\citep{epley2016mechanics}. Ideologies might be more appealing if they make a person feel important, but they also have to make sense---if only to the person and his fellow ideologues.

Another answer is that people adopt ideologies for purely instrumental reasons. This answer suggests, for example, that a neoliberal ideology~\citep{slobodian2018globalists} will attract wealthy capitalists seeking lower taxes. Ideologies are not ``single issue'' positions, however, and it is it not clear if a person can forecast whether or not adopting an ideology will lead to the desired outcome. Nor are ideologies necessarily easy to abandon, and the capitalist who becomes a neoliberal because of taxes may later find himself arguing against government subsidies that would be in his own self-interest. \cite{fanon2008black} provides a particularly affecting account of how ideologies can turn upon those who adopt them.

Second, how are ideologies learned? Does one dimension, explanatory or behavioral, tend to precede the other? \cite{althusser2014reproduction} cites Blaise Pascal as saying ``kneel down, move your lips in prayer, and you will believe''---\emph{i.e.}, in our framing, that the behavioral precedes the explanatory---but it is also somewhat hard to see how something as sophisticated as as world-view could be transmitted solely through imitation. Humans spend a great deal of time providing reasons to each other~\citep{mercier2017enigma}, and it seems equally plausible that a person comes to adopt an ideology by first being shown how it explains things that matter to them.

Our work aims to provide new answers to both questions, using the Red Pill's extremist ideology as a case study. As we will show, automated methods allow us to extract the explanatory and behavioral components of the Red Pill ideology in discussions online; we can then study the ways in which people engage with different components over time. This enables us to study what is attractive to a user at first, where such an initially intrigued user goes next, and what happens to those users if they choose to engage with the ideology over timescales of months and years. One of our innovations is the use of Hidden Markov Modeling to study the ideological formation of a user over time, which allows us to infer an underlying ``position'' in ideological space on the basis of the user's engagement with the surface content.

\section{Methods}

Forums associated with the Red Pill ideology are a core component of a larger ``manosphere''. Prior work has studied these groups as a source of extremism~\citep{jane2018systemic,van2018digesting}, antifeminism and misogyny~\citep{lin2017,mountford2018topic,farrell2019exploring,bhgt}, as a companion to in-person ``pickup culture''~\cite{o2018seduction}, and as virtual spaces that provide new affordances for articulating an aggrieved manhood~\citep{Masculinities2016Schmitz,ging2019alphas}. Researchers interested in the extreme views in these spaces focus on how such views are propagated through anonymity, accompanied by aggressive use of trolling and doxxing, and how rhetorical tactics, such as the strategic use of irony, help normalize extremist material under the guise of ambiguity and exacerbates group polarization~\citep{poe}.

Our primary data source here are the archives of the main ``subreddit'' group, r/TheRedPill (``r/TRP''). As with all of Reddit's content, r/TRP follows the familiar pattern of an online bulletin board. An r/TRP user is presented with a scrolling list of  ``submissions'', which range from extended 10,000 word user-authored essays, to laconic, one-sentence commentaries on a link to a news article hosted elsewhere. Users are able to comment on submissions, and to reply to the comments of others; they are also able to create new submissions (despite the name, these appear automatically). Both comments and submissions can be up-voted and down-voted by other users, and these votes influence placement and prominence. For simplicity, below, we use the word ``post'' to refer to both submissions and comments.

Users are free to delete anything they contribute, as are privileged users (``moderators'', or ``mods'') who intervene to remove posts they consider inappropriate. They are also able to edit the ``sidebar'', on the right-hand of the screen, that all users are presented with; the sidebar contains links to the site's norms and rules, and a curated list of submissions intended to serve as an introduction to the group. We used the pushshift.io website, which continually scrapes and archives content from Reddit, to gather both comments and submissions for further analysis. Our full download collected 3,091,290 comments and 115,527 submissions from 26 October 2012, r/TRP's founding date, to 25 January 2021. Our sample is nearly complete; due to what appears to be a database error in the scraped archive, we are missing some posts from the first three months of 2019.

\subsection{Topic Modelling}

In order to determine the semantic content of the site, we build a topic model~\citep{tm} from the user comments and submissions. A topic model is an unsupervised clustering method that takes a collection of documents (in this case, all of the individual comments and submissions), and decomposes them into sparse combinations of ``topics'', or co-occurring word patterns. Topic models are a well-tested and long-standing tool that provides a window into the semantics of the content; a vast array of studies in different domains has used them to show how texts combine together different ideas. Each topic that the topic model discovers corresponds to a pattern of word usage, and the output of the model includes both a description of the word patterns, and a decomposition of each comment and submission into a weighted combination of these patterns. A key advantage of topic modeling is that it is unsupervised: the method infers both the word patterns and the decompositions simultaneously.

In order to build the topic model, we pre-process the raw text by filtering stop-words and finding common 2-grams, and then use the MALLET software package~\citep{McCallumMALLET} for the inference itself. Because our goal is to study the full spectrum of user generated content, we model both comments and submissions together. For our final analysis, we use a 100-topic model. We inspect the output of this model by hand, and drop a total of seventeen ``non-semantic'' topics associated with automatically-posted material, complaints about the site or users, idiosyncratic slang, and meta discussion (thanking other users, asking for links, and so forth); our analysis focuses on the remaining 83. The Online Appendix lists each of these topics individually, showing the top fifty words in the topic, and providing twenty examples of texts heavily loaded on the topic.

Following \cite{gabe}, we then compute the \emph{linkage} between topics. Linkage measures the extent to which two topics tend to co-appear; two topics (say, $i$ and $j$) have high linkage when, if you encounter topic $i$, you are more likely to also encounter topic $j$. In the simplest case, we measure ``text-level'' linkage, \emph{i.e.}, the extent to which the appearance of topic $i$ in a particular text predicts the appears of topic $j$. We first compute the joint probability, $p_{ij}$ of seeing topic $i$ and topic $j$ in texts drawn from a particular era; if $p_i(k)$ is the fraction of topic $i$ found in text $k$, we have
\begin{equation}
p_{ij} = \frac{1}{N} \sum_{k=1}^N p_i(k) p_j(k),
\label{joint}
\end{equation}
and then linkage is defined as
\begin{equation}
R_{ij} = \log_2{\frac{p_{ij}}{p_i p_j}}.
\label{linkage}
\end{equation} 
Linkage is connected to the information-theoretic concept of mutual information; indeed, mutual information is simply the system's overall average linkage, weighted by the joint probability distribution $p_{ij}$. Text-level linkage provides a fine-grained map of how ideas are connected together in any particular post or comment. We can also measure linkage at the level of users; in Eq.~\ref{joint}, we can set $p_i(k)$ to the probability that user $k$ uses topic $i$ over all their posts, and sum is over all users, rather than all texts. User-level linkage tells us the extent to which, if a user tends to write about topic $i$, they become more likely to write about topic $j$; in our analysis, we find that both networks are broadly similar, meaning that individual users tend to associate topics in the same fashion as the group as a whole, and so, for simplicity, we present only the text-level results below.

Linkage enables us to do two things. First, it is a measure of the cognitive structure of the ideas in play; if two topics have high linkage, this is a signal that there is something that repeatedly draws people to associate the underlying ideas. Second, it provides a powerful tool for coarse-graining the topics found by a topic model. When the topics are represented as a nodes in a network, with the edge weights given by the linkage strength, network clustering algorithms such as Louvain clustering~\citep{blondel2008fast} can reveal large collections of co-associated topics that correspond to distinct conceptual units.

A topic model requires that we specify, ahead of time, the number of topics to find. This corresponds to choosing an effective ``resolution''---\emph{i.e.}, with more topics one can resolve, or split, word patterns that would be merged in a model with fewer topics. Ideally, one would like to choose that number on the basis of a statistical model-selection criterion. Somewhat surprisingly, however, the standard criteria are difficult to estimate well~\citep{eval} and can provide sub-optimal answers~\citep{tea}, and best practice in NLP places a premium on manual analysis and interpretation to validate parameter settings~\citep{183611}. 

In our analysis we followed work such as~\cite{allen2017topic} and  experimented with a range of different choices for the number of topics, between 20 and 100. We found that even at the highest-resolution setting of 100, the topic groupings were both stable on re-running the fits, and interpretable, meaning that (after filtering) any particular topic could be associated with a semantic theme, on the basis of both the top words in the topic, and an examination of strongly-loaded posts. These are shown in the Online Appendix. 

Because our linkage clustering method ends up grouping large numbers of topics into the same cluster, our results are largely insensitive to the detailed parameter choices of the topic model. We find that the results of our linkage model (discussed below; the grouping into clusters and the overall themes of each cluster) not only match manual inspection, but do not require fine tuning of parameters. If a lower-resolution model would merge two topics, these are expected to have high linkage already; similarly, if a higher-resolution model would split a topic into two, the resulting pair is expected to have high linkage. In either case, the merge or the split occurs within a cluster, and the overall cluster structure does not change very much. 

\subsection{Longitudinal Sample Construction}

A goal of our analysis is to study users in a longitudinal fashion as they engage with, and potentially learn, the group's ideas. The simplest approach to this requires a list of ``clean'' users who have no previous interactions with groups in the manosphere. Using the list of manosphere groups provided by~\cite{ribeiro2020evolution}, we construct a list of users who, (1) have not previously interacted with any manosphere groups on that list; but also, (2) have at least one post to a non-manosphere group, at the time they first appear in our data.

Criterion (1) means that we are not including users who have already learned the ideology from interaction with other groups in the manosphere; this provides the cleanest possible set of ``newcomers''. Criterion (2) is useful because it enables us to filter out ``special purpose'' accounts---\emph{i.e.}, accounts created by a Reddit user with the express purpose of interacting with manosphere content, and nothing else. Special purpose accounts (sometimes called ``throwaways'') are occasionally used on Reddit by individuals who want to make sure that their regular accounts---whose posting histories may include personally identifying information---are not tied to potentially inflammatory content, and may be a sign that the user is not as unfamiliar with r/TRP as one might otherwise assume. 

Producing a ``clean'' user subset is a critical step in the analysis because it helps eliminate users who may have already accumulated significant experience with the ideology elsewhere. This gives us a clearer insight into what happens immediately after first contact. Such as list is not perfect---in particular, it doesn't account for the ways in which a user might interact with these ideas in other fora, or the extent to which they only read the material before deciding to join the discussion---but it provides a first step for selecting on the initially less aware and less committed.

Our clean user set contains 39,838 users, all of whom have their first post on r/TRP somewhere between the beginning of 2016 and the end of 2019. By comparison, the set of special purpose accounts is much smaller, 6741; and the total number of users in this time period is 61,404---\emph{i.e.}, roughly a quarter of the participants on r/TRP during that period began posting on the Red Pill only after having had some interaction with other parts of the manosphere, a finding consistent with that of \cite{ribeiro2020evolution}'s study of cross-posting.

\subsection{Modeling Individual-Level Behavior using Hidden Markov Modeling}

Each user in our clean sample is associated with a list of posts they have made. A particular user with no prior record of engagement with the manosphere might have entered r/TRP at (say) January 1st, 2016, and made five different posts over the course of six months, with the last recorded post on June 1st. Our topic model will have allocated those five posts to different weighted combinations of the 100 topics in the data. This provides a trace of the user's interests over the course of their engagement with the site, and, in particular, with the kind of material they are generating for others on the site. 

A Hidden Markov Model (HMM) provides a way to model the dynamics of such a user, in terms of a shifting, unobserved ``hidden'' state. At any particular point in time, the user's hidden state corresponds to a preference for different topics in the system, and the dominant topic of the user's next post is draw from a distribution over those topics. Once the user makes the post, the hidden state then updates; the user might transition to a different hidden state, corresponding to a different set of preferences, or remain in the same one.

A HMM provides a model of an underlying set of user interests, with their own temporal logic, that express themselves in observed posting behaviors. A crucial feature of an HMM is that observed behavior of the user at any isolated point in time does not completely dictate what he will do next; there may be more than one hidden state consistent with the kind of post they made. Intuitively, this captures how different underlying ideological preferences can express themselves, at times, in the same fashion. 

The inference of that hidden state is key to correctly modeling a user's ideology. Consider (to take a general example) two users on a politics forum, both of whom express an opposition to racism. User A's expression of an opposition to racism comes from an attachment to a free-market and libertarian ideology, while User B's expression comes from an attachment to a socialist ideology. These users will have, generically, both different pasts, and different futures, and if we want to understand their distinct ideological trajectories, we will need to attempt to infer---on the basis of their other remarks---that their expressed opposition comes from very different conceptual frameworks. This is precisely what the HMM does, in the context of a particularly simple, robust framework that can be expressed as a generative Bayesian model. The inference itself is done using SFIHMM~\citep{dedeo2016conflict}, which implements the standard expectation-maximization algorithm \citep{baum1966statistical}, along with a model-selection method (Akaike Information Criterion, AIC) that automatically selects the optimal number of states in a Bayesian fashion.

\section{Results}

We first present the results of our topic modeling, which allows us to characterize semantics the Red Pill ideology. We then present the results on population structure and ideological formation, based on our individual-level modeling of the ``clean'' user subsample.

\subsection{Topic Modelling and Linkage Network Construction} 

\begin{figure}
\includegraphics[width=0.95\linewidth]{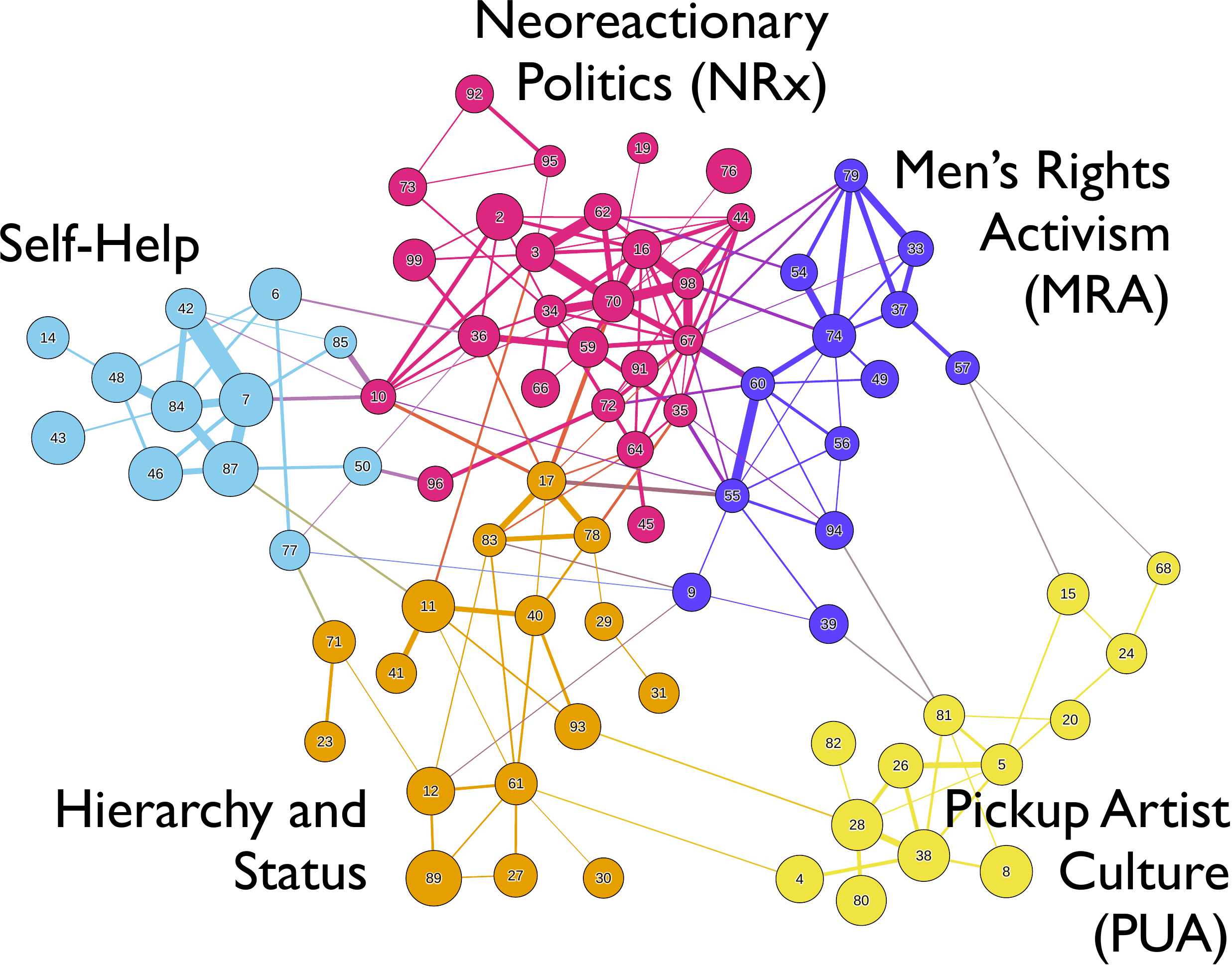}
\caption{The semantic structure of posts on r/TRP. Each node in the network corresponds to a topic (persistent word pattern) found in posts, with node size proportional to the fraction of attention devoted to the topic overall. An edge between two nodes indicates that the two corresponding topics are often found together in the same post. For example, Topic 17 (pseudo-evolutionary psychology; the orange node roughly in the center of the diagram) is often found in posts with Topic 70 (``decline of civilization''), Topic 83 (``hypergamy'' and evolutionary drivers behind male status experiences), and Topic 55 (parental rights). Nodes are arranged so that high-linkage topics are drawn closer together visually; we then use the Louvain clustering algorithm to detect groups of highly interlinked nodes, which are colored by group. Table~\ref{cluster_table} provides details on each of the clusters. \label{text_level}}
\end{figure}

The results of our topic modelling most easily visualized using the linkage network in Fig.~\ref{text_level}. This provides a graphical representation of how users create connections between different ideas; a high linkage (strong edge) between two topics means that the presence of one topic in a comment or submission is predictive of the other. For example, when a post includes talk about lifting weights as a form of self-improvement (Topic 87), it often also includes discussion of nutrition and testosterone (Topic 7), or martial arts and physical violence (Topic 50); see Appendix~\ref{topics_list} for a complete account of each topic. 

Having represented the topics in this fashion, we can use the Louvain clustering algorithm to reveal the large-scale structure. It finds five principal clusters, listed in Table~\ref{cluster_table}.  The two clusters, ``self-help'' and ``pickup-artist culture'' are primarily behavioral and concern the ways in which one ought to behave. Two clusters, ``hierarchy and status'', and ``neoreactionary politics'' are explanatory: they are focused in providing users with unifying accounts of the way the world works. A fifth cluster, ``men's rights activism'', is a mixture of the two dimensions.

\begin{table}
\begin{tabularx}{\textwidth}{l|X}
Cluster & Selections from Comments \\  \hline
Self-help & ``Prior to my morning meditation I have a cold shower and drink a coffee whilst filling out The Daily Stoic and making my today's plan.'' (Topic 84) \\
& ``Also stop watching porn for fucks sake. You can't improve if you're watching another man fuck the girl that you want. It's as simple as that. Sex is natural, porn is artificial.'' (Topic 48) \\
&  ``I guess from my perspective, if the depression and loss of interest is so bad, then there is a root cause or more that need to be identified and dealt with. ... This is all from personal experience. Last summer I hit a very low point in my life.'' (Topic 14) \\  \hline
PUA & ``Only approach women who give you a bunch of IOI's [indicators of interest]. Bonus points if she floats around in your vicinity to make your approach easier. At this point its all about having the balls to talk to her.'' (Topic 28) \\ 
& ``For instance in night game if a girl I meet in a public space (a club or pub) doesn't want to kiss me the night I meet her, unless there's a very specific logistics reason for her not to (like unable to isolate), I don't even bother getting her number.'' (Topic 38) \\ 
& ``Was plating this girl for about 5 months. Everything was fine as is, all of sudden I sense that she is losing interest; no longer initiated communication, told me twice in one month that she was on her period, long replies to text...'' (Topic 8) \\ \hline \hline
MRA & ``Women shouldn't have careers. They shouldn't get ``compensation'' for being a ``stay at home mother'' AKA mother. Career mothers practice child abuse.'' (Topic 55) \\ 
& ``Western spousal support laws are thinly veiled ``give women money'' laws, as any man who tries to use them rapidly finds out.'' (Topic 60) \\
& ``You can't deduct contributions to an IRA if you make over a certain amount so, most high income earners opt for the Roth which isn't tax deductible but grows tax free'' (Topic 74) \\ \hline \hline
Status & ``Harsh climate's prioritize alphas, the men who were able to hunt. An extreme example would be the tribes living in polar regions, where meat is the only food source and women cannot forage for berries and small game, thus making it advantageous for them to ally with the best alpha, the best hunter.'' (Topic 17) \\
& ``Alpha men embrace their malehood, the sexual strategies that maximize their genetic potential. Beta males play the provider game and resist their polygynous nature in an attempt to lure a female with his provision only.'' (Topic 83) \\ 
& ``So even if you are a short guy who builds a ton of muscle you will still be low status if when people look at you they know you are a bitch mentally. You can see it in their eyes \& mannerisms.'' (Topic 11) \\ \hline
NRx & ``In every example, these cultures began to rise when women were required to be virgins at marriage and to be monogamous for life. All of these cultures began to decline when women were given rights, were not required to be virgins at marriage, when divorce was common, and marriage was in decline.'' (Topic 70) \\ 
& ``You know: destroy the family unit by injecting Marxist class warfare into sexual dynamics (e.g. women are the stand-in proletariat fighting for equal wages with the supposedly bourgeois men). Gender Studies is Socialism of the Sexes, and its just as full of corrupt bureaucrats carefully toeing the party line.'' (Topic 16) \\ 
& ``I personally think that white people had their reasons for their prejudice. Fact: Black people didn't build a sustainable culture long before the white man set foot on their land. Look at Detroit, which is mostly black.'' (Topic 3) \\ 
\end{tabularx}
\vspace{0.3cm}
\caption{Examples of topics drawn from the five major clusters of the linkage network. The Self-help and PUA clusters are primarily behavioral, and focus on the ways in which men ought to live their lives and how they ought to relate to others. Meanwhile, the Status and NRx clusters are more explanatory, focusing on argument-making and providing general and unifying accounts of many different phenomena. The MRA cluster is a mixture of behavioral and explanatory. \label{cluster_table}}
\end{table}

The {\bf self-help} cluster includes themes such as emotional self-control and stoicism (Topic 6), breaking addictions, including to pornography (Topic 48), and mental health (Topic 14), along with themes related to exercise and nutrition. The {\bf pickup-artist culture} (PUA) cluster includes discussion of ``game'' and how to approach women in bars and social contexts (Topic 28), on communicating in early stages of dating (Topic 38), and advice on body language and non-verbal communication (Topic 80).  Broadly speaking, both of these clusters lie along the behavioral dimension of the Red Pill ideology. They are concerned with the details of how participants behave towards others and how they discipline their own bodies and minds to as to achieve right action. Taken together, they are about 28\% of all posts on the site.

While the PUA cluster is strongly focused on the social contexts of young men, the third cluster is associated with an earlier, predecessor ideology, {\bf Men's Rights Activism} (MRA). It includes discussion of divorce law and alimony (Topic 60), single mothers and paternity (Topic 55), as well as subjects commonly associated with the MGTOW movement such as achieving financial freedom (Topics 74). The MRA cluster is the smallest in our data, consisting of only 7\% of all posts.

The fourth cluster, ``{\bf hierarchy and status}'', is the largest in the system by total posts. It contains abstract and (pseudo) scientific discussion of evolutionary psychology (``pseudo-evo-psych''; Topic 17), the social hierarchy of ``alpha'' and ``beta'' males (Topic 78), and discussion of the sexual ``strategies'' that women use to attain high-status partners (Topic 83). It also contains more specific discussion of how to rate oneself, other men, and potential female sexual partners on the basis of (for example), height and body shape (Topic 11), visual appearance (Topic 41), or social status and wealth (Topic 40). Roughly 47\% of all posts are dominated by patterns from the status cluster.

Finally, the fifth cluster contains ideas associated with {\bf neoreactionary politics} (NRx), a recently-identified political ideology connected to the contemporary alt-right movement~\citep{jones2019neoreactionary,lyons,sandifer2018neoreaction}. Dominant themes in the NRx cluster include the idea that women, as a group, are responsible for the decline of civilization (Topic 70), and that feminism is a collectivist ideology spread by a conspiracy of elites (Topic 16), as well as discussion of contemporary U.S. politics, including racial politics (Topic 3) and comparisons between different countries (Topic 62). NRx accounts for roughly 19\% of all posts.

The Status and NRx clusters lie along the explanatory, sense-making dimension of the Red Pill ideology. Posts weighted on these word patterns tend to provide synthetic and systematic explanations for the way things are. They may make scientific arguments, or political arguments, that show how a number of distinct pieces of evidence connect together. NRx topics include explanations in terms of conspiracy theories (\emph{i.e.}, that attempt to explain major world events and features of society in terms of the secret actions of a malevolent group), while Status topics tend to focus on explanations in terms of a pseudo-scientific version of evolutionary psychology. The MRA cluster appears to be a mixture of the two. Some of the content is behavioral, and provides explicit instruction on how to manage finances, divorce, and parenting. Other content is explanatory, and tries to explain (for example) how the different ways that men feel oppression is driven by anti-male prejudice in society and the law.

\subsection{Population Structure}

\begin{figure}
\includegraphics[width=0.95\linewidth]{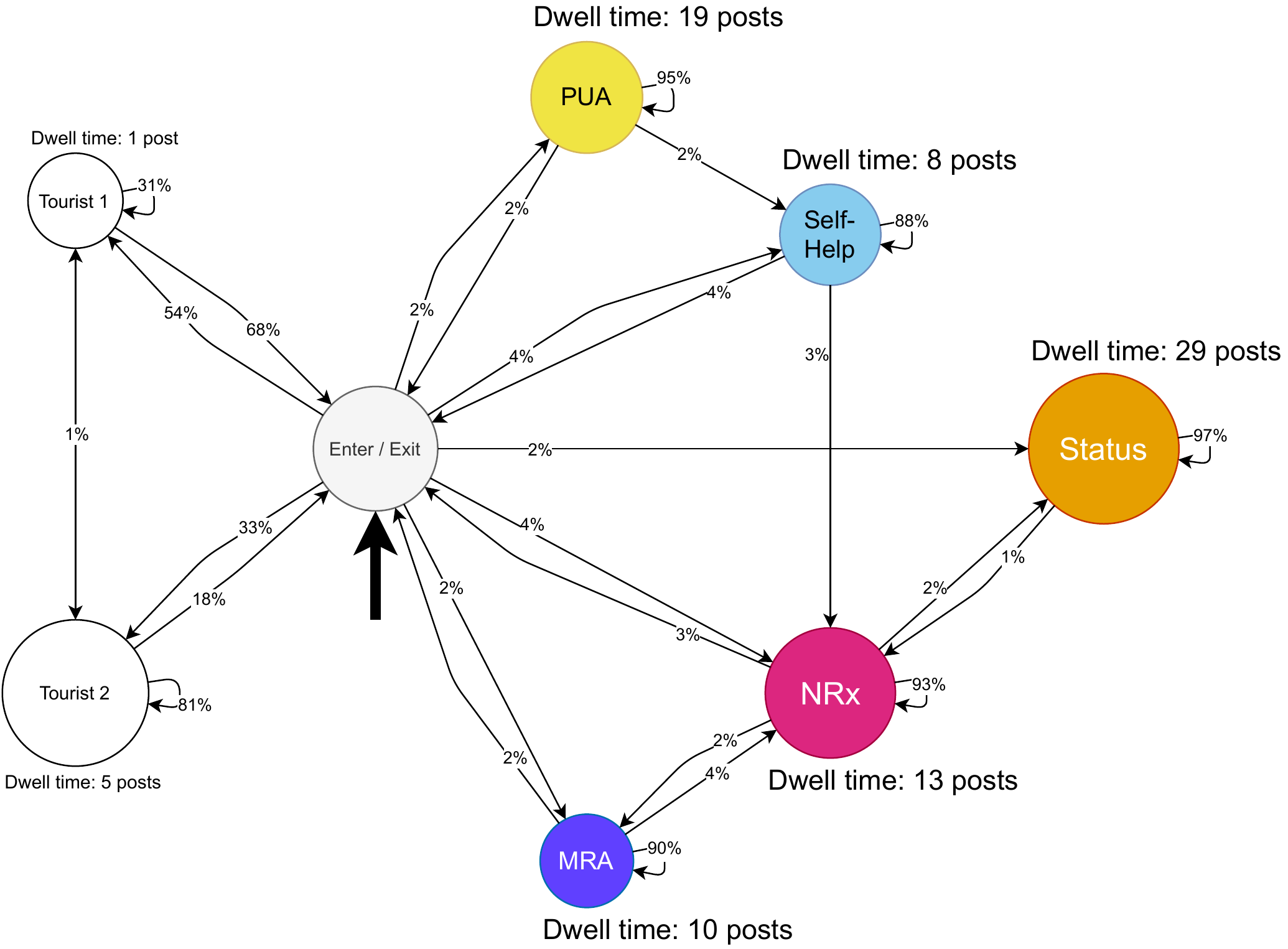}
\caption{Ideological formation online: the HMM for users from our longitudinal ``clean'' sample. A user starts at the ``enter/exit'' state, and then proceeds, with different probabilities, to one of seven different content patterns with different dwell times. Two patterns, labelled ``tourist'', are associated with very quick exits from the system; for example, 54\% of users who enter the system post content in the ``tourist 1'' pattern, and leave after, on average, just one post. Five styles, however, are associated with much longer-term residence. The ``NRx'' content pattern, for example, is far stickier; users will remain in the NRx posting style for an average of 13 posts before either exiting, or going on to (for example) post content associated with Men's Rights Activism, or Status. States are labelled according to the results of Table~\ref{types}, \emph{e.g.}, the user state associated with (significantly) above average posts from the self-help cluster is labelled ``self-help''.  For visual clarity in this diagram, very weak transition probabilities are not shown. \label{hmm}}
\end{figure}

\begin{table}
\begin{tabular}{l|l|l|l|l|l}
 & Self-Help & PUA & Status & MRA & NRx \\ \hline
Self-Help  State & {\bf +190\%} & -29\% & -33\% & -47\% & -36\% \\ 
PUA  State & -10\% & {\bf +107\%} & +11\% & -8\% & -35\% \\ 
Status  State & -24\% & -4\% & {\bf +48\%} & -15\% & -7\% \\ 
MRA State & -30\% & -25\% & -25\% & {\bf+179\%} & -17\% \\ 
NRx  State & -49\% & -49\% & -37\% & -33\% & {\bf +62\%} \\ 
\end{tabular}
\vspace{0.3cm}
\caption{The relationship between resident type and posting content. For each state of the HMM (\emph{e.g.}, ``Self-Help State''), we show the probability of a post dominated by the semantics of the five clusters, relative to baseline. Broadly speaking, at any point in time, a resident user is predisposed towards one of the five semantic clusters shown in Fig.~\ref{text_level}; \emph{e.g.}, a user in what we call the ``self-help state'' is 190\% more likely to post content related to self-help, and 47\% less likely to post content related to Men's Rights Activism. \label{types}}
\end{table}

A visualization of the recovered Hidden Markov Model for user evolution is shown in Fig.~\ref{hmm}. The analysis reveals the existence of two distinct types of users, which we refer to as ``tourists'' and ``residents''. Tourists have low overall engagement with the system; they make only a few posts before leaving, in one of two possible posting patterns that matches the overall distribution of semantics on the system. The residents fall into five user types with distinct patterns of engagement; as shown in Table~\ref{types}, these roughly align with different parts of the linkage network.

\begin{figure}
\includegraphics[width=0.8\linewidth]{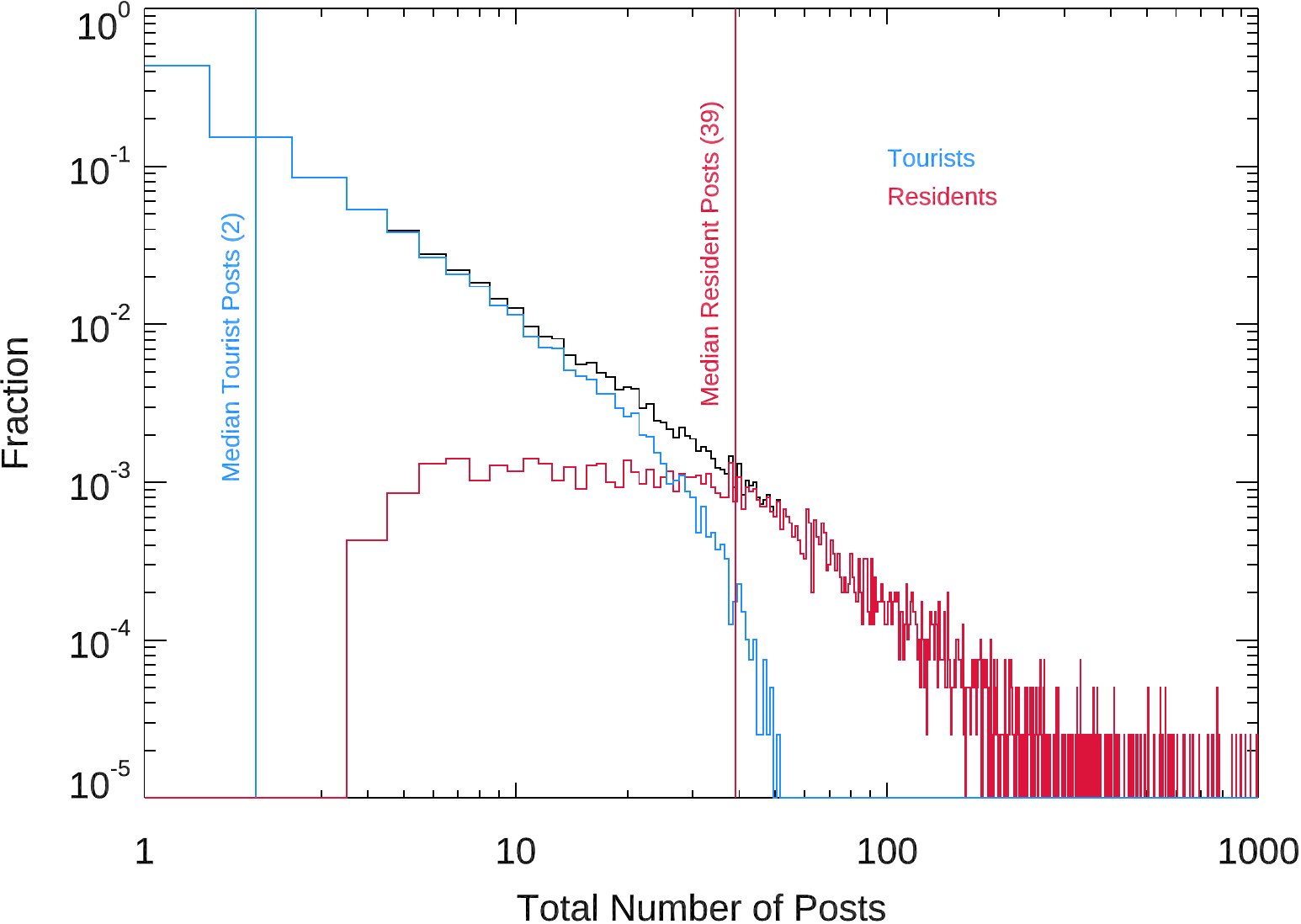}
\caption{Distribution of total numbers of posts for the tourist (blue) and resident (red) user types. Users who take the right-hand branch on Fig.~\ref{hmm} remain nearly twenty times longer (by median number of posts) compared to the ``tourists''. \label{num}}
\end{figure}
The HMM reveals a ``tourist--resident' structure, where most visitors (87\%) are ``tourists'' who have only casual interactions with the group; their posting patterns are consistent with sampling the content on the site at random and choosing one particular aspect of the content to imitate. Meanwhile, the remaining 13\% are ``residents'', who generate the majority of the content (62\% of all posts, and 65\% of all words). This can be seen in both the numbers of posts the different user types make; Fig.~\ref{num}. The vast majority of posters who stay longer than a few posts entered the system on the right-hand branch of Fig.~\ref{hmm}.

\subsection{Ideological Formation}

The HMM of Fig.~\ref{hmm} allows us to distinguish different modes of engagement with Red Pill ideology. It also allows us to determine which aspects of that ideology are ``appealing'' (\emph{i.e.}, characteristic of the initial engagement of a long-term resident) and which are ``sticky'' (\emph{i.e.}, characteristic of later behavior); as shown in Table~\ref{trans}, the most obvious pattern is a shift from the ``appealing'', behavioral content associated with self-help talk, and towards the ``sticky'' explanatory content of the status pattern.

\begin{table}[h!]
\begin{tabular}{l|l|l|l|l|l}
State & Fraction & Fraction & Dwell Time & Residence & Next State \\ 
 & (First) & (Overall) & (Posts) & (Months) & \\ \hline 
Self-Help & 30\% & 14\% & 8 & 7.9 & Exit (38\%), NRx (23\%), PUA (18\%) \\ 
PUA & 18\% & 18\% & 19 &  11 & Exit (36\%), Self-Help (36\%), MRA (17\%) \\ 
MRA & 13\% & 12\% & 10 & 7.8 & NRx (37\%), Exit (21\%), Status (16\%) \\
Status & 12\% & 32\% & 29 & 14 & NRx (34\%), Exit (26\%), MRA (19\%) \\ 
NRx & 27\% & 24\% & 13 & 8.1 & Exit (39\%), MRA (21\%), Status (20\%) \\ 
\end{tabular}
\vspace{0.3cm}
\caption{Relationships between the different resident patterns. For example, 30\% of residents are expected to begin in the self-help state; those in that state spend roughly eight posts there before, say, shifting over to NRx (23\% of the time), PUA (18\% of the time), or leaving the system entirely (38\% of the time). While the plurality of residents join the group as self-help users in the behavioral mode, those who stay either began with, or shift over to, more explanatory content, most notably ``Status'' and ``NRx''. ``Residence'' here is defined as the (median) observed total amount of time spent in the system by a user primarily associated with that state; for example, a user that posts primarily in the ``self-help'' pattern has a median residence time on the site of 7.9 months. \label{trans}}
\end{table}

The HMM also allows us to understand ideological formation, \emph{i.e.}, how users learn. Consider, for example, a user who begins in the self-help pattern. A user who remains in this pattern will (38\% of the time) leave the system, after roughly eight posts; however, 23\% of the time, the user will ``advance'' from the self-help pattern to the NRx pattern, where they will remain for roughly ten posts. At this point, they might exit (39\% of the time), or continue on to either the MRA pattern (21\% of the time), or the Status pattern (20\% of the time).

In practice, these shifts have major impacts on user trajectories. For example, the median residence time of a user that begins in the ``self-help'' state is 7.9 months. However, if a user that begins in the self-help state shifts over to the explanatory NRx state, their median residence time nearly doubles, to 11.4 months. Adopting the Status pattern has the largest effect size of all; those who switch to the Status state increase their median residence time from 7.9 months to 18.5 months. We do not see a strong effect in the other direction; for example, a user that begins in the Status state has a median residence time of 14.0 months; if they switch to the self-help state, the median residence time rises, by a much smaller amount, to 14.8 months. The fact that ``self-help, then status'' has a longer residence time than ``status, then self-help'' points to a more complex psychological process where behavioral shifts can make the individual more susceptible to later explanatory indoctrination.

\section{Discussion}


Seen from the outside, the posts on an extremist forum can look like a confusing mix of in-group language and deliberately offensive, politically motivated trolls~\citep{nagle2017kill}. The first outcome of our work is to show how simple tools can bring order to this apparent chaos. Given a topic model where word patterns serve as a proxy for the semantics of the underlying ideas, the information theoretic concept of linkage can then reveal how different pieces of an ideology connect together into larger conceptual units. These results, in turn, provide support for a general ``two-dimensional'' account of ideology as a combination of distinct explanatory and behavioral modules. 

In the case of the Red Pill ideology, for example, some clusters we uncover are concerned with how members of the group are supposed act in the world, including (in the self-help cluster) their relationship to pornography use, eating, emotional regulation, and physical exercise, and (in the PUA cluster) how they ought to judge women as potential partners, and how to treat them when seeking romantic relationships. Other parts of the ideology are concerned with providing explanations for the way things are. These include the ideas about genetic differences between the races, theories about the role of women in the history of civilizations, and the existence of elite conspiracies that censor and conceal unpleasant truths (in the NRx cluster), the existences of biological imperatives, hardcoded into evolution, that force women to act in certain ways, and that make some men more worthy than others (in the Status cluster).

The second outcome of our work concerns how newcomers interact with these groups. A simple model of individual-level behavior provides a rough division of newcomers into two populations, tourists and residents, with the first group sampling randomly and leaving quickly, and the second group showing more sustained and systematic engagement, with distinct preferences for different features of the ideology. This two-population model has important implications for the experience of the group itself: nine times out of ten, content is produced by a user who has had little interaction with the group, and will almost certainly leave soon after. Many of the same features that make online communities sources of concern for those tracking radicalization---low barriers to entry, the use of pseudonyms rather than real names, proximity to more ``ordinary'' content---also appear to flood these systems with the disengaged.

The third outcome concerns how the newcomers who do persist navigate these ideas. Table~\ref{trans} provides the clearest insight into what attracts individuals in the short term, versus what leads them to persist and integrate further into the community. Our major result here is a contrast between, on the one hand, the wider appeal of the ideology's behavioral dimension, and, on the other, the power of explanatory content to draw users more deeply in. We suggest that a key feature of an ideology, and, at least in this first case study, the aspect that induces long-term and in-depth engagement, is what appeals to the ``sense-making'' drive~\citep{CHATER2016137,motiv}.

In addition to helping us understand the relationship between behavioral and explanatory aspects of ideology, these results also help us understand the particular nature of extremist ideologies surrounding sex and gender. The central role played by the ``Status'' cluster, both in the overall semantics of the system and in how it serves to attract and retain users, confirms a key insight of work such as \cite{ging2019alphas}: that Red Pill ideology is concerned not just with the superiority of men over women, but also with explaining, and defining, the gradations in status and power between men. Similar points have also been made by \cite{manne} in her analysis of a prominent self-help book associated with the movement, and \cite{o2018seduction}'s fieldwork in (offline) ``seduction'' groups has noted the particular role that pseudo-evo-psych plays in justification and explanation. 

Status may not just be what keeps individuals within the ideology. It may also be what draws them towards violence. Status concerns are often, as both \cite{manne} and \cite{masc} note, significant motivating factors for young men who go on to shooting rampages; they also appear to be a basic preoccupation of men who engage in domestic violence~\citep{jukes2020men}. As \cite{quest} argue, the violence associated with extremist ideology is motivated by a ``quest for personal significance''; it may well be that social status serves this role in violence associated with sexist ideology. In this case, the status cluster seen in Fig.~\ref{text_level} connects the ideology to a theory of personal significance that could then serve as a motivation for violence~\citep{more_quest}.

The linkage network of Fig.~\ref{text_level} also points to an intimate connection between sexist ideology and political theories associated with white supremacy and anti-Semitism, which confirms recent work by non-profit organizations on the intersection between these emergent sexist ideologies and older, more familiar traditions~\citep{adl,icct}. Indeed, misogyny itself---as an outcome of a ``particular'' sexist ideology such as that found on the Red Pill, or as a consequence of more ordinary forms of sexism---may be one of the links that connects extremist ideologies to political violence \citep{nyt}.

A key question, for both scientists and policy makers, is the ways in which people are drawn in to extremist groups, and this work provides at least a partial answer. It also suggests that the same individuals may leave these groups when they become disenchanted by the explanations. This may be driven in part by engagement with individuals explicitly adopt a critical stance against the ideology; large-scale studies of these kinds of ``counter-speech'' may help us understand the underlying psychology of explanatory disenchantment and the different forms of argument-making that lead to it~\citep{garland2020impact}. There are many reasons why individuals will become disillusioned with an extremist ideology, and the process is a complex one particularly when it includes relationships in the offline world (see, \emph{e.g.}, \citealp{leaving}). In a world of online interactions, at least, explanatory disenchantment may play a central role in disillusionment and exit.


\section*{Acknowledgements}

We thank Hahrie Hahn, Henry Farrell, Kathleen Blee, Suresh Naidu, Joshua Garland, and others for helpful conversations.

\clearpage

\appendix
\section{Appendix: Topic Model Samples}
\label{topics_list}

A key step in our analysis is the interpretation of the output of our topic model in terms of meaningful semantics. There are two different ways to do this. The first method is to look at the characteristic words associated with each topic; often this provides a clear idea of the general themes. The second method is to look at documents heavily weighted on that topic. This can provide important additional context, in part because it can reveal what different words in the topic are doing when they combine together. 

For example, consider Topic 2, the first semantic topic on our list below (recall that we drop 17 of the topics because they are associated with non-semantic word patterns such as automated bots). Topic 2's word list contains a number of argument-related words (``argument'', ``point'', ``wrong'', ``logic'', etc). By visiting a few of the sample posts, it becomes clear that it tracks a certain set of reasoning norms. For example, the full text of the third example is 

\begin{quotation}\noindent 
``You are just choosing to believe whatever you want. Saying they don't know is the same as not having evidence to support it. Science is pretty straightforward, either you can prove it or you can't. Everything in your last comment is just speculation, you can speculate and it's OK, what is not ok is that you present your hypothesis as a fact, it isn't a fact. You are using a misinterpretation of a scientific article about flies to support your philosophical ideas. You are entitled to an opinion and beliefs, just don't mislead other people into it.''
\end{quotation}

Our sample below contains posts from the beginning of 2018 and 2019, that have at least 20 words (after filtering for stop words) and that are weighted at least 50\% on the topic of interest. For each post, we present the fractions of the next two highly weighted (semantic) topics, which can help in interpretation. In a few cases, the comment may no longer be accessible via the hyperlink; this is because users and moderators are allowed to delete comments, but they can be caught in the pushshift.io database before this happens, and thus enter our analysis.

\subsection{Topic 2 (Nrx)}

\noindent [\emph{n.b.}, Topic 2 is provided as a sample; remaining topic data available at \url{http://santafe.edu/~simon/pd_online_appendix.pdf}]

{\bf Top words}:

 {\tt argument point wrong said logic saying say evidence trying arguments fact argue debate logical anything opinion arguing nothing post discussion statement points agree facts someone making claim prove comment made based something actually without talking true reason disagree right believe instead simply stupid using understand bullshit try completely proof use claims response valid explain clearly reasoning person rational correct anyone mean makes rather attack statements false topic position fallacy case actual conclusion opinions conversation science either see idea support never calling personal theory theres example matter irrelevant ideas hominem sound scientific criticism exactly internet seem read issue entire assumptions could thought original attempt back thinking whether convince yet obvious come otherwise cannot obviously also lack stated counter defend yes truth call idiot nonsense subject thread words emotional clear ignorant assumption provide ridiculous really says win intelligent proven therefore view done logically stating attacks debating please assume insults pointing enough waste need shaming sense intellectual pointless premise discuss literally address seriously absolutely insult engage look conclusions examples strawman context ignore justify respond attacking anecdotal comments sure reply ops question reasonable incorrect discussing real whole troll dumb dismiss claiming simple retarded show seems information ignorance lets take must fine things basis straw 
}

 {\bf Sample posts}: 

\url{http://reddit.com/r/TheRedPill/comments/74a5ia/red_piller_here_whos_been_working_in_strip_clubs/dsc5ihh/}; 49 filtered words.  Topic 2 (50\%); Topic 70 (17\%); Topic 73 (6\%)

\url{http://reddit.com/r/TheRedPill/comments/74a5ia/red_piller_here_whos_been_working_in_strip_clubs/dsc5j7u/}; 49 filtered words.  Topic 2 (50\%); Topic 70 (19\%); Topic 19 (6\%)

\url{http://reddit.com/r/TheRedPill/comments/7owq63/its_not_your_girl_its_just_your_turn_married/dsftc98/}; 31 filtered words.  Topic 2 (67\%); Topic 10 (8\%); Topic 61 (3\%)

\url{http://reddit.com/r/TheRedPill/comments/7pzss2/the_culmination_of_the_prowomen_movements_and/dsncwgm/}; 48 filtered words.  Topic 2 (52\%); Topic 8 (14\%); Topic 56 (4\%)

\url{http://reddit.com/r/TheRedPill/comments/7r0t92/made_this_comment_on_an_aziz_ansari_thread_in_2x/dsu1lq7/}; 21 filtered words.  Topic 2 (54\%); Topic 94 (16\%); Topic 49 (4\%)

\url{http://reddit.com/r/TheRedPill/comments/7sf3pq/jordan_petersons_cathy_newman_interview_good/dt4umj7/}; 24 filtered words.  Topic 2 (50\%); Topic 92 (11\%); Topic 66 (7\%)

\url{http://reddit.com/r/TheRedPill/comments/7ts74u/pet_theories_and_newunfinished_ideas_megathread/dth2xjb/}; 27 filtered words.  Topic 2 (60\%); Topic 77 (13\%); Topic 35 (6\%)

\url{http://reddit.com/r/TheRedPill/comments/7v4afp/jordan_peterson_is_not_your_friend/dtrfmfw/}; 22 filtered words.  Topic 2 (52\%); Topic 9 (11\%); Topic 53 (11\%)

\url{http://reddit.com/r/TheRedPill/comments/7y88js/toxic_argumentative_females_should_be_avoided_at/duf1r3n/}; 31 filtered words.  Topic 2 (58\%); Topic 45 (20\%); Topic 38 (3\%)

\url{http://reddit.com/r/TheRedPill/comments/7yug4s/focussing_on_pussy_isnt_beta_behavior/dujpd2t/}; 24 filtered words.  Topic 2 (64\%); Topic 5 (4\%); Topic 50 (4\%)

\url{http://reddit.com/r/TheRedPill/comments/bcjtns/women_in_ground_combat/ekzwgx6/}; 33 filtered words.  Topic 2 (58\%); Topic 80 (8\%); Topic 56 (3\%)

\url{http://reddit.com/r/TheRedPill/comments/bi9uic/alpha_fucks_beta_bucks_according_to_a/em2oq8y/}; 36 filtered words.  Topic 2 (51\%); Topic 64 (10\%); Topic 7 (5\%)

\url{http://reddit.com/r/TheRedPill/comments/bsa48u/why_you_should_never_get_married/eonceul/}; 24 filtered words.  Topic 2 (52\%); Topic 76 (11\%); Topic 83 (4\%)

\url{http://reddit.com/r/TheRedPill/comments/bsa48u/why_you_should_never_get_married/eont340/}; 80 filtered words.  Topic 2 (60\%); Topic 60 (7\%); Topic 36 (5\%)

\url{http://reddit.com/r/TheRedPill/comments/c06ve7/what_is_the_red_pill/er52mw4/}; 22 filtered words.  Topic 2 (56\%); Topic 51 (11\%); Topic 57 (8\%)

\url{http://reddit.com/r/TheRedPill/comments/c28gpx/the_evolutionary_perspective_on_fake_orgasms_is/erj42cw/}; 79 filtered words.  Topic 2 (62\%); Topic 42 (10\%); Topic 64 (8\%)

\url{http://reddit.com/r/TheRedPill/comments/c28gpx/the_evolutionary_perspective_on_fake_orgasms_is/erj9jm1/}; 23 filtered words.  Topic 2 (72\%); Topic 17 (11\%); Topic 4 (0\%)

\url{http://reddit.com/r/TheRedPill/comments/eonaqw/dont_get_cucked_by_cuckcervatives/fee7pnp/}; 24 filtered words.  Topic 2 (52\%); Topic 94 (14\%); Topic 73 (14\%)

\url{http://reddit.com/r/TheRedPill/comments/eonaqw/dont_get_cucked_by_cuckcervatives/feei2y8/}; 42 filtered words.  Topic 2 (51\%); Topic 16 (11\%); Topic 31 (9\%)

\url{http://reddit.com/r/TheRedPill/comments/gqr8yz/go_macros_kill_micros_take_some_real_pills/frw9m9l/}; 24 filtered words.  Topic 2 (59\%); Topic 82 (7\%); Topic 93 (7\%)

\clearpage

\bibliographystyle{abbrvnat} 
\bibliography{preprint}

\begin{thebibliography}{64}
\providecommand{\natexlab}[1]{#1}
\providecommand{\url}[1]{\texttt{#1}}
\expandafter\ifx\csname urlstyle\endcsname\relax
  \providecommand{\doi}[1]{doi: #1}\else
  \providecommand{\doi}{doi: \begingroup \urlstyle{rm}\Url}\fi

\bibitem[Aikin(2013)]{poe}
S.~F. Aikin.
\newblock Poe's law, group polarization, and argumentative failure in religious
  and political discourse.
\newblock \emph{Social Semiotics}, 23\penalty0 (3):\penalty0 301--317, 2013.
\newblock \doi{10.1080/10350330.2012.719728}.

\bibitem[Allen et~al.(2017)Allen, Luo, Murdock, Pu, Wang, Zhai, and
  Zhao]{allen2017topic}
C.~Allen, H.~Luo, J.~Murdock, J.~Pu, X.~Wang, Y.~Zhai, and K.~Zhao.
\newblock Topic modeling the h{\`a}n di{\u{a}}n ancient classics.
\newblock \emph{Journal of Cultural Analytics}, 2017.
\newblock \doi{10.22148/001c.11882}.

\bibitem[Althusser(1971)]{althusser2014reproduction}
L.~Althusser.
\newblock \emph{On the reproduction of capitalism: Ideology and ideological
  state apparatuses}.
\newblock Verso, 1971.
\newblock Revised edition 2014. Translated by G.\ M.\ Goshgarian.

\bibitem[Baum and Petrie(1966)]{baum1966statistical}
L.~E. Baum and T.~Petrie.
\newblock Statistical inference for probabilistic functions of finite state
  {M}arkov chains.
\newblock \emph{The Annals of Mathematical Statistics}, 37\penalty0
  (6):\penalty0 1554--1563, 1966.

\bibitem[Blei and Lafferty(2009)]{tm}
D.~M. Blei and J.~Lafferty.
\newblock Topic models.
\newblock In A.~N. Srivastava and M.~Sahami, editors, \emph{Text Mining:
  Classification, Clustering, and Applications}. CRC Press, 2009.

\bibitem[Blondel et~al.(2008)Blondel, Guillaume, Lambiotte, and
  Lefebvre]{blondel2008fast}
V.~D. Blondel, J.-L. Guillaume, R.~Lambiotte, and E.~Lefebvre.
\newblock Fast unfolding of communities in large networks.
\newblock \emph{Journal of Statistical Mechanics: Theory and Experiment},
  2008\penalty0 (10):\penalty0 P10008, 2008.

\bibitem[Bosman et~al.(2019)Bosman, Taylor, and Arango]{nyt}
J.~Bosman, K.~Taylor, and T.~Arango.
\newblock A common trait among mass killers: Hatred toward women.
\newblock \emph{New York Times}, Aug. 2019.
\newblock
  \url{https://www.nytimes.com/2019/08/10/us/mass-shootings-misogyny-dayton.html}.
  Last accessed 19 March 2021.

\bibitem[Bourdieu(1990)]{bourdieu1990logic}
P.~Bourdieu.
\newblock \emph{The Logic of Practice}.
\newblock Stanford University Press, 1990.

\bibitem[Braly et~al.(2018)Braly, Parent, and DeLucia]{braly2018threats}
A.~M. Braly, M.~C. Parent, and P.~R. DeLucia.
\newblock Do threats to masculinity result in more aggressive driving behavior?
\newblock \emph{Psychology of Men \& Masculinity}, 19\penalty0 (4):\penalty0
  540, 2018.

\bibitem[Chang et~al.(2009)Chang, Boyd-Graber, Gerrish, Wang, and Blei]{tea}
J.~Chang, J.~Boyd-Graber, S.~Gerrish, C.~Wang, and D.~M. Blei.
\newblock Reading tea leaves: How humans interpret topic models.
\newblock In \emph{Proceedings of the 22nd International Conference on Neural
  Information Processing Systems}, NIPS'09, pages 288--296, Red Hook, NY, USA,
  2009. Curran Associates Inc.
\newblock ISBN 9781615679119.

\bibitem[Chater and Loewenstein(2016)]{CHATER2016137}
N.~Chater and G.~Loewenstein.
\newblock The under-appreciated drive for sense-making.
\newblock \emph{Journal of Economic Behavior \& Organization}, 126:\penalty0
  137--154, 2016.
\newblock ISSN 0167-2681.
\newblock \doi{https://doi.org/10.1016/j.jebo.2015.10.016}.
\newblock URL
  \url{https://www.sciencedirect.com/science/article/pii/S0167268115002838}.
\newblock Thriving through Balance.

\bibitem[Chwe(2013)]{chwe2013rational}
M.~S. Chwe.
\newblock \emph{Rational Ritual: Culture, Coordination, and Common Knowledge}.
\newblock Princeton University Press, 2013.
\newblock ISBN 9780691158280.

\bibitem[Connor et~al.(2016)Connor, Glick, and Fiske]{fiske}
R.~A. Connor, P.~Glick, and S.~T. Fiske.
\newblock Ambivalent sexism in the twenty-first century.
\newblock In C.~G. Sibley and F.~K. Barlow, editors, \emph{The Cambridge
  Handbook of the Psychology of Prejudice}, pages 295--320. Cambridge
  University Press, 2016.

\bibitem[Converse(1964)]{converse}
P.~E. Converse.
\newblock The nature of belief systems in mass publics.
\newblock In D.~E. Apter, editor, \emph{Ideology and its discontents}. The Free
  Press of Glencoe, New York, 1964.

\bibitem[DeDeo(2016)]{dedeo2016conflict}
S.~DeDeo.
\newblock Conflict and computation on {W}ikipedia: A finite-state machine
  analysis of editor interactions.
\newblock \emph{Future Internet}, 8\penalty0 (3):\penalty0 31, 2016.

\bibitem[DiBranco(2020)]{icct}
A.~DiBranco.
\newblock Male supremacist terrorism as a rising threat.
\newblock \emph{International Centre for Counter-Terrorism Perspectives}, 2020.
\newblock Available at
  \url{https://icct.nl/publication/male-supremacist-terrorism-as-a-rising-threat/}.
  Last accessed 19 March 2021.

\bibitem[Elias(1939)]{elias2000civilizing}
N.~Elias.
\newblock \emph{The Civilizing Process: Sociogenetic and Psychogenetic
  Investigations, Revised Edition}.
\newblock Wiley, 1939.
\newblock ISBN 9780631221609.
\newblock Revised edition published 2000. Translated by Edmund Jephcott.

\bibitem[Epley and Gilovich(2016)]{epley2016mechanics}
N.~Epley and T.~Gilovich.
\newblock The mechanics of motivated reasoning.
\newblock \emph{{Journal of Economic Perspectives}}, 30\penalty0 (3):\penalty0
  133--40, 2016.

\bibitem[Fanon(1967)]{fanon2008black}
F.~Fanon.
\newblock \emph{Black Skin, White Masks}.
\newblock Grove Press, 1967.
\newblock Revised edition 2008. Translated by R. Philcox.

\bibitem[Farrell et~al.(2019)Farrell, Fernandez, Novotny, and
  Alani]{farrell2019exploring}
T.~Farrell, M.~Fernandez, J.~Novotny, and H.~Alani.
\newblock Exploring misogyny across the manosphere in reddit.
\newblock In \emph{Proceedings of the 10th ACM Conference on Web Science},
  pages 87--96, 2019.

\bibitem[Friedrich(1965)]{friedrich1965ideology}
C.~J. Friedrich.
\newblock Ideology in politics: A theoretical comment.
\newblock \emph{Slavic Review}, 24\penalty0 (4):\penalty0 612--616, 1965.

\bibitem[Garland et~al.(2020)Garland, Ghazi-Zahedi, Young, H{'e}bert-Dufresne,
  and Galesic]{garland2020impact}
J.~Garland, K.~Ghazi-Zahedi, J.-G. Young, L.~H{'e}bert-Dufresne, and
  M.~Galesic.
\newblock Impact and dynamics of hate and counter speech online, 2020.
\newblock arXiv cs.SI: 2009.08392.

\bibitem[Gerdes et~al.(2018)Gerdes, Alto, Jadaszewski, D'Auria, and
  Levant]{gerdes2018content}
Z.~T. Gerdes, K.~M. Alto, S.~Jadaszewski, F.~D'Auria, and R.~F. Levant.
\newblock {A content analysis of research on masculinity ideologies using all
  forms of the Male Role Norms Inventory (MRNI)}.
\newblock \emph{Psychology of Men \& Masculinity}, 19\penalty0 (4):\penalty0
  584, 2018.

\bibitem[Ging(2017)]{ging2019alphas}
D.~Ging.
\newblock Alphas, betas, and incels: Theorizing the masculinities of the
  manosphere.
\newblock \emph{Men and Masculinities}, 22\penalty0 (4):\penalty0 638--657,
  2017.

\bibitem[Glick et~al.(1997)Glick, Diebold, Bailey-Werner, and
  Zhu]{glick1997two}
P.~Glick, J.~Diebold, B.~Bailey-Werner, and L.~Zhu.
\newblock The two faces of {A}dam: Ambivalent sexism and polarized attitudes
  toward women.
\newblock \emph{Personality and Social Psychology Bulletin}, 23\penalty0
  (12):\penalty0 1323--1334, 1997.

\bibitem[Graham et~al.(2013)Graham, Haidt, Koleva, Motyl, Iyer, Wojcik, and
  Ditto]{GRAHAM201355}
J.~Graham, J.~Haidt, S.~Koleva, M.~Motyl, R.~Iyer, S.~P. Wojcik, and P.~H.
  Ditto.
\newblock Moral foundations theory: The pragmatic validity of moral pluralism.
\newblock In P.~Devine and A.~Plant, editors, \emph{Advances in Experimental
  Social Psychology}, volume~47, pages 55--130. Academic Press, 2013.
\newblock \doi{https://doi.org/10.1016/B978-0-12-407236-7.00002-4}.

\bibitem[Jane(2018)]{jane2018systemic}
E.~A. Jane.
\newblock Systemic misogyny exposed: Translating rapeglish from the manosphere
  with a random rape threat generator.
\newblock \emph{{International Journal of Cultural Studies}}, 21\penalty0
  (6):\penalty0 661--680, 2018.

\bibitem[Jones(2019)]{jones2019neoreactionary}
A.~Jones.
\newblock From neoreactionary theory to the alt-right.
\newblock In \emph{Critical Theory and the Humanities in the Age of the
  Alt-Right}, pages 101--120. Springer, 2019.

\bibitem[Jukes(1999)]{jukes2020men}
A.~E. Jukes.
\newblock \emph{Men Who Batter Women}.
\newblock Routledge, 1999.

\bibitem[Kalish and Kimmel(2010)]{masc}
R.~Kalish and M.~Kimmel.
\newblock Suicide by mass murder: Masculinity, aggrieved entitlement, and
  rampage school shootings.
\newblock \emph{Health Sociology Review}, 19\penalty0 (4):\penalty0 451--464,
  2010.
\newblock \doi{10.5172/hesr.2010.19.4.451}.

\bibitem[Kruglanski et~al.(2014)Kruglanski, Gelfand, Bélanger, Sheveland,
  Hetiarachchi, and Gunaratna]{more_quest}
A.~W. Kruglanski, M.~J. Gelfand, J.~J. Bélanger, A.~Sheveland,
  M.~Hetiarachchi, and R.~Gunaratna.
\newblock The psychology of radicalization and deradicalization: How
  significance quest impacts violent extremism.
\newblock \emph{Political Psychology}, 35\penalty0 (S1):\penalty0 69--93, 2014.
\newblock \doi{10.1111/pops.12163}.

\bibitem[Kruglanski et~al.(2019)Kruglanski, Fernandez, Factor, and
  Szumowska]{quest}
A.~W. Kruglanski, J.~R. Fernandez, A.~R. Factor, and E.~Szumowska.
\newblock Cognitive mechanisms in violent extremism.
\newblock \emph{Cognition}, 188:\penalty0 116--123, 2019.
\newblock ISSN 0010-0277.
\newblock \doi{10.1016/j.cognition.2018.11.008}.

\bibitem[Latif et~al.(2020)Latif, Blee, DeMichele, Simi, and
  Alexander]{leaving}
M.~Latif, K.~Blee, M.~DeMichele, P.~Simi, and S.~Alexander.
\newblock {Why White Supremacist Women Become Disillusioned, and Why They
  Leave}.
\newblock \emph{The Sociological Quarterly}, 61\penalty0 (3):\penalty0
  367--388, 2020.
\newblock \doi{10.1080/00380253.2019.1625733}.

\bibitem[League(2018)]{adl}
A.-D. League.
\newblock When women are the enemy: The intersection of misogyny and white
  supremacy, 2018.
\newblock Available at \url{https://www.adl.org/media/11707/download}. Last
  accessed 19 March 2021.

\bibitem[Levant et~al.(1992)Levant, Hirsch, Celentano, Cozza, Hill, MacEachern,
  Schnedeker, et~al.]{levant1992male}
R.~Levant, L.~Hirsch, E.~Celentano, T.~Cozza, S.~Hill, M.~MacEachern,
  J.~Schnedeker, et~al.
\newblock The male role: An investigation of norms and stereotypes.
\newblock \emph{Journal of Mental Health Counseling}, 14\penalty0 (3):\penalty0
  325--337, 1992.

\bibitem[Lin(2017)]{lin2017}
J.~L. Lin.
\newblock {Antifeminism Online: MGTOW (Men Going Their Own Way)}.
\newblock In S.~F. Urte Undine~Fr{\"o}mming, Steffen~K{\"o}hn and M.~Terry,
  editors, \emph{Digital Environments: Ethnographic Perspectives Across Global
  Online and Offline Spaces}, pages 77--96. Transcript Verlag, Bielefeld,
  Germany, 2017.
\newblock \doi{10.25969/mediarep/2891}.

\bibitem[Lyons(2019)]{lyons}
M.~N. Lyons.
\newblock {CTRL-ALT-DELETE}: the origins and ideology of the {Alternative
  Right}.
\newblock \emph{Political Research Associates}, 2019.

\bibitem[Mandaville(2010)]{mandaville2010global}
P.~Mandaville.
\newblock \emph{Global Political Islam}.
\newblock Taylor \& Francis, 2010.
\newblock ISBN 9781134341368.

\bibitem[Manne(2017)]{manne2017down}
K.~Manne.
\newblock \emph{Down girl: The logic of misogyny}.
\newblock Oxford University Press, 2017.

\bibitem[Manne(2018)]{manne}
K.~Manne.
\newblock Reconsider the lobster.
\newblock \emph{Times Literary Supplement}, 5 2018.

\bibitem[Marx and Engels(1845)]{marx}
K.~Marx and F.~Engels.
\newblock \emph{The German Ideology}.
\newblock Lawrence \& Wishart, 1845.
\newblock Updated version edited by C. J. Arthur, 1970.

\bibitem[McCallum(2002)]{McCallumMALLET}
A.~K. McCallum.
\newblock {MALLET}: A machine learning for language toolkit.
\newblock 2002.
\newblock URL \url{http://mallet.cs.umass.edu}.

\bibitem[Mercier and Sperber(2017)]{mercier2017enigma}
H.~Mercier and D.~Sperber.
\newblock \emph{The Enigma of Reason}.
\newblock Harvard University Press, 2017.

\bibitem[Mountford(2018)]{mountford2018topic}
J.~Mountford.
\newblock Topic modeling the red pill.
\newblock \emph{Social Sciences}, 7\penalty0 (3):\penalty0 42, 2018.

\bibitem[Nagle(2017)]{nagle2017kill}
A.~Nagle.
\newblock \emph{{Kill All Normies: Online Culture Wars From 4Chan And Tumblr To
  Trump And The Alt-Right}}.
\newblock John Hunt Publishing, 2017.
\newblock ISBN 9781785355448.

\bibitem[O'Neill(2018)]{o2018seduction}
R.~O'Neill.
\newblock \emph{Seduction: Men, masculinity and mediated intimacy}.
\newblock John Wiley \& Sons, New York, NY, 2018.

\bibitem[Oswald et~al.(2019)Oswald, Baalbaki, and Kirkman]{self_esteem}
D.~L. Oswald, M.~Baalbaki, and M.~Kirkman.
\newblock Experiences with benevolent sexism: Scale development and
  associations with women's well-being.
\newblock \emph{Sex Roles}, 80\penalty0 (5):\penalty0 362--380, 2019.

\bibitem[Ribeiro et~al.(2020)Ribeiro, Blackburn, Bradlyn, Cristofaro,
  Stringhini, Long, Greenberg, and Zannettou]{ribeiro2020evolution}
M.~H. Ribeiro, J.~Blackburn, B.~Bradlyn, E.~D. Cristofaro, G.~Stringhini,
  S.~Long, S.~Greenberg, and S.~Zannettou.
\newblock The evolution of the manosphere across the web, 2020.

\bibitem[Roberts et~al.(2016)Roberts, Stewart, and Tingley]{183611}
M.~Roberts, B.~Stewart, and D.~Tingley.
\newblock \emph{Navigating the Local Modes of Big Data: The Case of Topic
  Models}.
\newblock Cambridge University Press, New York, NY, USA, 2016.

\bibitem[Salmon et~al.(2021)Salmon, Thompson, and DeDeo]{gabe}
G.~L. Salmon, W.~H.~W. Thompson, and S.~DeDeo.
\newblock {Consilience and the cultural evolution of conceptual networks in
  London's Royal Society}.
\newblock 2021.
\newblock Preprint.

\bibitem[Sandifer and Graham(2018)]{sandifer2018neoreaction}
E.~Sandifer and J.~Graham.
\newblock \emph{{Neoreaction a Basilisk: Essays on and Around the Alt-Right}}.
\newblock CreateSpace, 2018.
\newblock ISBN 9781986913997.

\bibitem[Schmitz and Kazyak(2016)]{Masculinities2016Schmitz}
R.~Schmitz and E.~Kazyak.
\newblock Masculinities in cyberspace: An analysis of portrayals of manhood in
  men's rights activist websites.
\newblock \emph{Social Sciences, 5(2), 18.}, page 1097184X18816118, 2016.
\newblock URL \url{http://dx.doi.org/10.3390/socsci5020018}.

\bibitem[Seliger(1970)]{seliger1970fundamental}
M.~Seliger.
\newblock Fundamental and operative ideology: The two principal dimensions of
  political argumentation.
\newblock \emph{Policy Sciences}, 1\penalty0 (1):\penalty0 325--338, 1970.

\bibitem[Seliger(1976)]{seliger2019ideology}
M.~Seliger.
\newblock \emph{Ideology and politics}.
\newblock Routledge, 1976.
\newblock Republished 2019.

\bibitem[Seliger(1979)]{seliger1979inseparability}
M.~Seliger.
\newblock The inseparability of ideology and politics: The case for a
  re-evaluation of the concept ``ideology''.
\newblock \emph{Revue Europ{\'e}enne des sciences sociales}, 17\penalty0
  (46):\penalty0 115--138, 1979.

\bibitem[Slobodian(2018)]{slobodian2018globalists}
Q.~Slobodian.
\newblock \emph{Globalists}.
\newblock Harvard University Press, 2018.

\bibitem[Talton(2020)]{bhgt}
W.~Talton.
\newblock \emph{{Sex, Gender, Sexual Assault, and Rape on The Red Pill: A
  Thematic and Linguistic Analysis}}.
\newblock Master's thesis, University of Central Florida, 2020.
\newblock URL \url{https://purls.library.ucf.edu/go/DP0023605}.

\bibitem[Thompson and Pleck(1995)]{pleck95}
E.~H. Thompson, Jr. and J.~H. Pleck.
\newblock Masculinity ideologies: A review of research instrumentation on men
  and masculinities.
\newblock In R.~F. Levant and W.~S. Pollack, editors, \emph{A new psychology of
  men}, pages 129---163. Basic Books, 1995.

\bibitem[Thompson(1987)]{thompson}
J.~B. Thompson.
\newblock Language and ideology: A framework for analysis.
\newblock \emph{The Sociological Review}, 35\penalty0 (3):\penalty0 516--536,
  1987.

\bibitem[Van~Valkenburgh(2018)]{van2018digesting}
S.~P. Van~Valkenburgh.
\newblock Digesting the red pill: Masculinity and neoliberalism in the
  manosphere.
\newblock \emph{{Men and Masculinities}}, page 1097184X18816118, 2018.

\bibitem[Wallach et~al.(2009)Wallach, Murray, Salakhutdinov, and Mimno]{eval}
H.~M. Wallach, I.~Murray, R.~Salakhutdinov, and D.~Mimno.
\newblock Evaluation methods for topic models.
\newblock In \emph{Proceedings of the 26th Annual International Conference on
  Machine Learning}, ICML '09, pages 1105--1112, New York, NY, USA, 2009.
  Association for Computing Machinery.
\newblock ISBN 9781605585161.
\newblock \doi{10.1145/1553374.1553515}.
\newblock URL \url{https://doi.org/10.1145/1553374.1553515}.

\bibitem[Wojtowicz and DeDeo(2020)]{WOJTOWICZ2020981}
Z.~Wojtowicz and S.~DeDeo.
\newblock From probability to consilience: How explanatory values implement
  bayesian reasoning.
\newblock \emph{Trends in Cognitive Sciences}, 24\penalty0 (12):\penalty0
  981--993, 2020.
\newblock ISSN 1364-6613.
\newblock \doi{https://doi.org/10.1016/j.tics.2020.09.013}.
\newblock URL
  \url{https://www.sciencedirect.com/science/article/pii/S136466132030228X}.

\bibitem[Wojtowicz et~al.(2021)Wojtowicz, Chater, and Loewenstein]{motiv}
Z.~Wojtowicz, N.~Chater, and G.~Loewenstein.
\newblock The motivational processes of sense-making.
\newblock In E.~Schulz, I.~Cogliati-Dezza, and C.~Wu, editors,
  \emph{Information-Seeking}. Cambridge University Press, Cambridge, UK, 2021.

\bibitem[Zanden(1959)]{zanden}
J.~W.~V. Zanden.
\newblock The ideology of white supremacy.
\newblock \emph{Journal of the History of Ideas}, 20\penalty0 (3):\penalty0
  385--402, 1959.
\newblock URL \url{http://www.jstor.org/stable/2708116}.

\end{thebibliography}


\end{document}